\documentclass{article}

\usepackage[utf8]{inputenc}
\usepackage[left=3cm,right=2cm,top=2cm,bottom=2cm]{geometry}
\usepackage[utf8]{inputenc}
\usepackage[T1]{fontenc}
\usepackage{bm}
\usepackage{xcolor}
\usepackage{graphicx}
\usepackage{url}
\usepackage{authblk}
\usepackage{amsmath,amssymb}

\DeclareMathAlphabet{\mathcal}{OMS}{cmsy}{m}{n}
\SetMathAlphabet{\mathcal}{bold}{OMS}{cmsy}{b}{n}

\newcommand{\eg}{{\it e.g.}, }
\newcommand{\ie}{{\it i.e.}, }
\newcommand{\MADAS}{\texttt{MADAS}}

\newcommand{\fsub}[1]{$_\mathrm{#1}$}

\title{\MADAS: A Python framework for assessing similarity in materials-science data}

\author[1, *]{Martin Kuban}
\author[1]{Santiago Rigamonti}
\author[1]{Claudia Draxl}
\affil[1]{Humboldt-Universit{\"a}t zu Berlin, Institut f\"ur Physik and CSMB, Berlin, 12489, Germany}

\affil[*]{corresponding author: Martin Kuban (kuban@physik.hu-berlin.de)}

\begin{document}

\maketitle

\begin{abstract}
Computational materials science produces large quantities of data, both in terms of high-throughput calculations and individual studies. Extracting knowledge from this large and heterogeneous pool of data is challenging due to the wide variety of computational methods and approximations, resulting in significant veracity in the sheer amount of available data. Here, we present \MADAS, a Python framework for computing similarity relations between material properties. It can be used to automate the download of data from various sources, compute descriptors and similarities between materials, analyze the relationship between materials through their properties, and can incorporate a variety of existing machine learning methods. We explain the design of the package and demonstrate its power with representative examples.

\end{abstract}

\section{Introduction}
\label{sec:Introduction}

The discovery of novel materials is a crucial aspect of technological progress. Therefore, much emphasis is placed on the in-depth characterization of materials as well as on the synthesis and prediction of new ones. As a result of such investigations, the community produces enormous amounts of data, including both experimental and computational results. In this context, the need to make data FAIR \cite{Wilkinson2016} ({\bf F}indable, {\bf A}ccessible, {\bf I}nteroperable, {\bf R}e-useable), has become evident, and large, publicly available databases have been created to store and retrieve these data. Such large data volumes come with new challenges, such as making data available and comprehensible for researchers from different communities.

\begin{figure}[!ht]
    \centering
    \includegraphics[width = 0.4\textwidth]{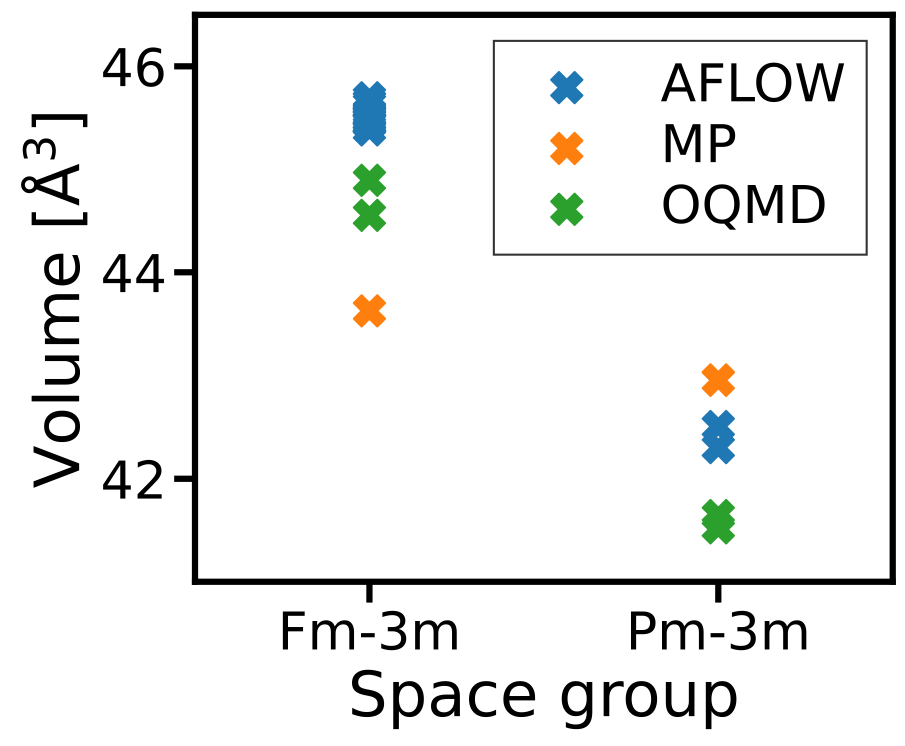}
    \caption{Comparison of the unit-cell volume of NaCl obtained by DFT across the databases AFLOW \cite{Curtarolo2012lib}, Materials Project\cite{Jain2013}, and OQMD\cite{Kirklin2015}. The code to reproduce this analysis can be found in this publication's GitHub repository (\protect \url{https://github.com/kubanmar/madas-examples}).}
    \label{fig:volume_comparison}
\end{figure}

At the same time, such data collections enable new types of analysis, employing data-analytics and machine-learning (ML) methods. Obviously, the more (reliable) data become available, the better the results are. Conversely, with increasing amount of data, quality control becomes a bottleneck. This problem seems to be less critical in large materials-science databases that contain results of high-throughput (HT) calculations, where a consistent set of parameters is used to simulate many --often several thousands-- of materials. In such HT efforts, the proper execution and convergence of the result is usually controlled by workflow description languages, such as ASR\cite{Gjerding2021} or Jobflow, or workflow engines, such as Fireworks\cite{Jain2015}, AiiDA\cite{Pizzi2016}, or MyQueue\cite{Mortensen2020}. However, data contained in different high-throughput databases may not be comparable because of different approximations and computational settings  used in the respective calculations. The effects of these differences are subject to recent studies, comparing all-electron to pseudopotential density-functional-theory codes\cite{Bosoni2023} by either using a dedicated benchmark dataset, or by comparing material properties contained in different databases directly. Here, we exemplify in Fig. \ref{fig:volume_comparison} the differences arising from different numerical approaches with results for two NaCl structures from three different HT materials databases. All calculations have been carried out with the VASP\cite{Kresse1996} code, employing density-functional theory (DFT) in the generalized-gradient approximation (GGA), more specifically the PBE parametrization. All structures are relaxed in terms of their volumes and atomic positions. We have verified that the structures are symmetrically equivalent by using the method described in Ref. \cite{Lonie2012} and implemented in ASE\cite{Larsen2017}, version v3.22.1. Besides the same DFT approach in all three cases, the reported volumes differ by up to 2 \AA$^3$. The results presented in Ref. \cite{Hegde2023} suggest that these discrepancies are due to differences in the plane-wave cutoff and different relaxation schemes. One may argue that these errors are comparatively small, however, given the simple structures, they are still significant and may lead to differences in the corresponding properties. Since high-quality {\it interoperable} data are an important prerequisite for achieving high-precision predictions by ML models, combining these data for such a task can lead to significant uncertainties. 

To quantify the uncertainty in data, appropriate metrics need to be defined. Similarity measures can help to assess the quality of heterogeneous data \cite{Kuban2022a}, discover trends and outliers \cite{Kuban2022b}, or generate maps of the contents of large materials databases\cite{Isayev2015}. During the last years, many \textit{descriptors} for different aspects of materials have been published, focusing, for instance, on the atomic\cite{Bartok2013, Huo2022, Langer2022, Himanen2020}  or electronic \cite{Isayev2015, Knosgaard2022, Kuban2022a} structure. Some of these descriptors, such as \textrm{SOAP} \cite{Bartok2013}, are specifically designed to measure the similarity between atomic configurations \cite{De2016}. The available descriptors largely vary in complexity, ranging from a single number to high-dimensional representations that may demand significant computational resources. Despite the existence of comprehensive libraries such as {\tt dscribe} \cite{Himanen2020} or {\tt matminer} \cite{Ward2018}, novel descriptors are usually published as stand-alone software packages. Integrating them in existing (or new) data-analysis workflows, requires 'glue code', that ensures the interoperability of the data formats required by the descriptors and other parts of the workflow. Such code tends to be application specific, not reusable, and hard to maintain. 

In this work, we present \MADAS, a Python framework that provides a modular, extendable, and simple interface to various tasks of similarity analysis of materials data. \MADAS\ has been previously used in various illustrative examples, including the search for similar materials\cite{Kuban2022b}, the analysis of the convergence behavior of the electronic structure in DFT calculations\cite{Kuban2022b}, and the clustering of materials based on the similarity in terms of their electronic density of states\cite{Kuban2022a}. The framework is equipped for various data-analysis tasks that use similarities with distinct but highly connected subtasks including (i) collection and storage of data from different sources, such as local file systems or remote databases, (ii) definition of material descriptors and similarity measures, (iii) calculation and storage of similarity relations, and (iv) analysis using a variety of techniques. In the following, we describe how we address related challenges by defining and implementing interfaces between the individual components of similarity analysis.

\section{Software components and their functionality}
\label{sec:components}
\begin{figure}[ht]
    \centering
    \includegraphics[width = 0.9\textwidth]{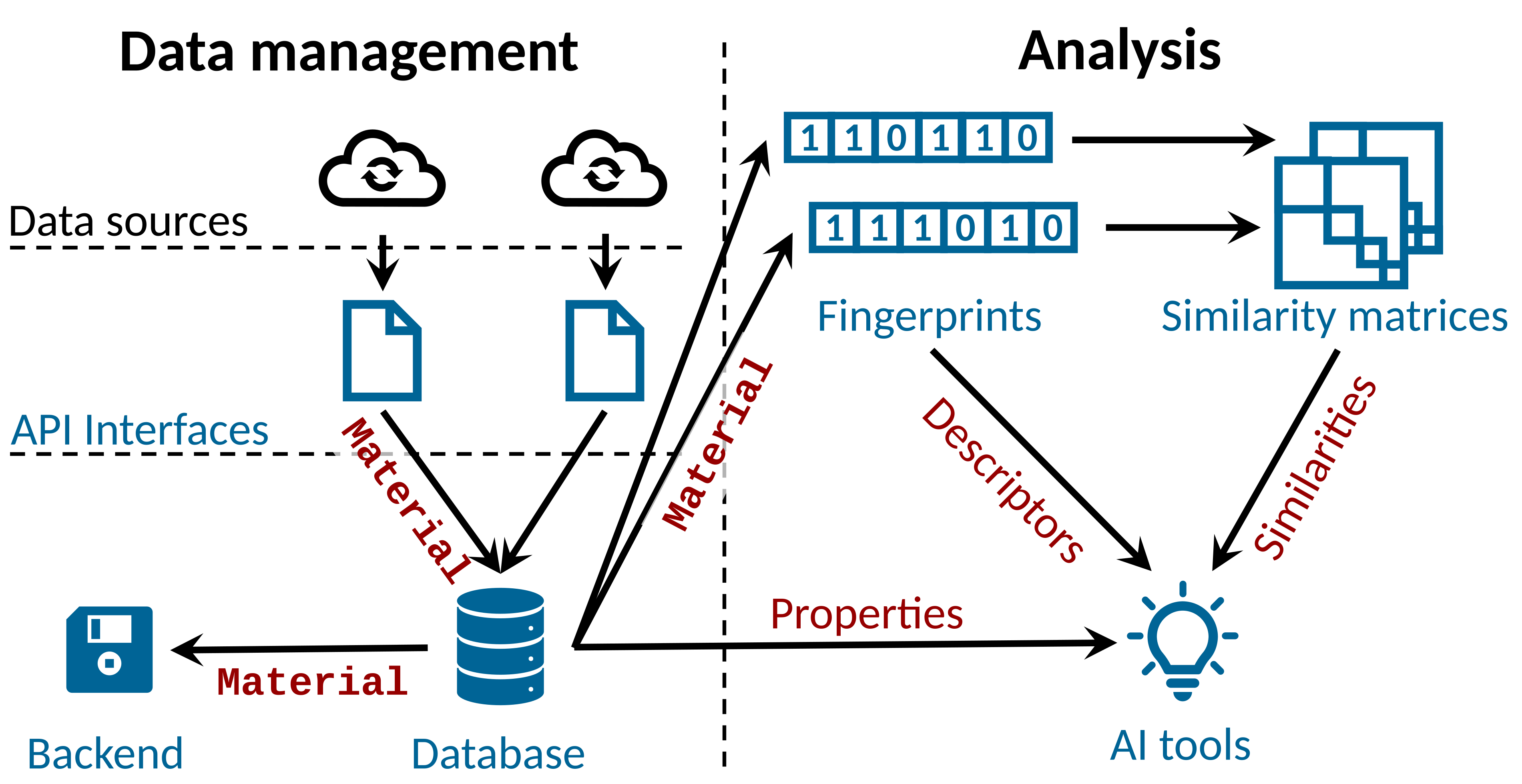}
    \caption{Schematic workflow diagram for using the \MADAS\ framework for data analysis. Symbols represent software components, arrows the data flow. Blue color indicates components that are explicitly implemented in \MADAS, red labels annotate the type of data that is exchanged. \texttt{Material} refers to the data class used within \MADAS.}
    \label{fig:workflow_diagram}
\end{figure}

An overview of the software architecture of \MADAS\ is shown in Fig. \ref{fig:workflow_diagram}. Data are exchanged between the components via a generic \texttt{Material} data class, which acts as a container for all data that can be used to calculate descriptors, a unique identifier (ID), and convenient methods to access these data. Below, the different components of \MADAS\ are described in detail, explaining their functions and providing use-cases.

\subsection{Collecting data from different sources}
To make use of databases, \eg for a data-analysis or ML task, users can decide for a (open source) database and download the data they want to use via web APIs that are maintained by the database provider. However, despite efforts towards API standards\cite{Andersen2021}, a providers' proprietary API may give access to more specific information. It may also be subject to frequent changes. Therefore, to obtain and use the rich data provided by online repositories, users are often required to write custom scripts. These are not necessarily published together with the scientific results. Often, a (reproducible) description of how the data were obtained is omitted altogether.

These challenges must be addressed from a practical point of view: Overall, users of online repositories have little influence on the design choices and availability of the data sources they rely on. This requires them to adapt their programs and workflows to any changes in online data. To efficiently work with the data under these conditions, programs must be easy to maintain and error-tolerant. To support users in this respect, \MADAS\ provides a Python class (called \texttt{APIClass}), which allows them to implement (and update) their own interfaces to external data (see Fig. \ref{fig:workflow_diagram}, top left). Equipped with  a common naming schema and data model, it can be used as a standardized template. That way, additional APIs can be added quickly. This template was used to write the code for generating Fig. \ref{fig:volume_comparison}, where minimal versions of API connections to the AFLOWlib \cite{Curtarolo2012lib}, Materials Project\cite{Jain2013}, and OQMD\cite{Kirklin2015} databases are utilized. By using the common naming schema and data class (\ie the \texttt{Material} class) for downloading data from different sources, neither the database, nor the data analysis pipeline (see Fig. \ref{fig:workflow_diagram}) need to be changed when new data are added or their descriptions are changed by the providers of external databases. Thus, data analysis workflows can be re-used, speeding up their development. Furthermore, \MADAS\ implements convenient error mitigation and logging mechanisms. It natively supports an interface to the NOMAD Archive\cite{Draxl2018}, built on top of the NOMAD web API\cite{Draxl2019}.

\subsection{Storing data locally} 
\label{sec:database}
The data contained in online repositories are often subject to changes, for instance when new calculations are added or numerical parameters are refined. To ensure a persistent set of data to work on and to avoid repeated downloads from the same source, it is necessary to store the data locally. Here, this is realized by using a database, the \texttt{MaterialsDatabase}. We note that, although the local storage of data can be realized by maintaining lists or binary object files, \eg supported by \texttt{numpy}\cite{Harris2020} or \texttt{pandas}\cite{McKinney2010}, during the course of a research project, the number of generated lists can become large, and therefore hard to maintain and update. Thus, using a database brings several benefits. First, the data are stored in a consistent way. Therefore, unless altered on purpose, the original input data are preserved and the reproducibility of the analysis is strongly supported. The \texttt{MaterialsDatabase} ensures uniqueness of the data w.r.t. unique identifiers (IDs), so-called \texttt{mid}s. With this persistent identification, no data are queried or stored twice, and the data entries can be linked with their original source. Second, the consistent usage of IDs allows for connecting different parts of \MADAS. For example, the results of clustering (see Sec. \ref{sec:AItools}) based on the electronic structure of materials, can be related to the respective atomic structures contained in the database. The actual storage of data is handled by a \textit{backend}, implemented in a \texttt{Backend} class, which is responsible for maintaining a connection to the database file, \ie for reading, writing, and updating its entries. By default, the \texttt{AtomsDatabase} of \texttt{ASE} \cite{Larsen2017} is used by the \texttt{Backend}, which implements a SQL schema for materials-science data. This enhances interoperability with other packages and allows for easy sharing of data. Different backends can be realized by writing custom \texttt{Backend} child classes that inherit its basic functionality. More details on how to achieve this, and tutorials for the implementation, can be found in the documentation (\url{https://madas.readthedocs.io}).

\subsection{Fingerprinting materials} 
\label{sec:fingerprinting}
The next step after data collection, is their analysis (right panel of Fig. \ref{fig:workflow_diagram}). For this, a suitable description of the data is required. One option is to extract properties directly from the database, \eg as tabular data. The \textit{MaterialsDatabase} implements this in its \textit{get\_property\_dataframe} method, which returns the data as a \texttt{pandas}\cite{McKinney2010} \texttt{DataFrame} object. Alternatively, \textit{fingerprints} can be used. In the context of this work, a fingerprint is the combination of a \textit{descriptor} and a \textit{similarity measure}. Descriptors (see also Sec. \ref{sec:Introduction}) are numerical representations of atomic configurations, and/or their respective properties. A similarity measure (sometimes also called a \textit{kernel}) is a function $S$ that maps any pair of descriptors $(A, B)$ to a similarity score $0 \leq S(A, B) \leq 1$. A similarity score of $S = 1$ ($S = 0$) means that the descriptors are completely identical (different). The choice of the function $S$ is arbitrary in general. However, depending on the problem that is addressed by fingerprinting materials, similarity measures with specific properties must be used. For most applications, symmetric measures, \ie $S(A, B) = S(B, A)$ are beneficial. For certain applications, such as clustering, a similarity measure whose complement ($1-S$) fulfills metric properties\cite{Willett1998} can be necessary (see, \eg Ref. \cite{Kuban2022a}). The verification of the former can be done analytically. Additionally, \MADAS\ provides a tool for verification of the metric properties for a given set of fingerprints, available as a Python class (\texttt{madas.analysis.MetricSpaceTest}).

We note that a large number of similarity measures that can be used are directly available in \texttt{scikit-learn}\cite{Pedregosa2011}. Additionally, any function $d \in [0,\infty)$, that assigns a distance to each pair of fingerprints, can be transformed to a similarity measure $S$ with $S = 1/(1+d)$.  

This combination of descriptor and similarity measure is represented in the data model of \MADAS: Each instance of a \texttt{Fingerprint} object is initialized with its respective similarity function. Then, the similarity between descriptors can be calculated by executing the \texttt{get\_similarity}-method of one of the fingerprints. When another similarity metric is required, it can be changed by calling the \texttt{set\_similarity\_function}-method, without changing other parts of the program.

An important distinction is to be made between fingerprint \textit{types} and \textit{parameterizations}. Fingerprints of the same \textit{type} use the same descriptor. Many implementations of descriptors, however, depend on specific choices of parameters, \eg a cutoff or the number of basis functions \cite{Bartok2013,Huo2022}. Descriptors obtained with different parameters cannot be compared in a meaningful way. 
To avoid this situation, \textit{Fingerprint} objects of the same type can be distinguished by their \textit{name}, which is an identifier that is representative of the parameterization. The method \texttt{get\_similarity} that is implemented for each \texttt{Fingerprint} checks that the type and name are the same for the fingerprints to be compared, such to ensure that the computation of similarity is meaningful.

The large variety of possible ways to compute material fingerprints results also in heterogeneous implementations that need to be adapted for every analysis workflow. Within \MADAS, this challenge is met by defining a generic \texttt{Fingerprint} class that is used consistently throughout all parts of the code. It allows for calculating the descriptors directly from database entries, \eg atomic structure and properties, and storing the results in the same database. This is particularly useful for reducing computational effort in cases where the calculation of the descriptor is resource intensive. The fingerprints can then be later reconstructed just from the database entries. This framework fosters rapid development of novel types of fingerprints, since they can be tested on real data, and the computed descriptors can be passed to the analysis pipeline to analyze the results. It also supports reuse of code, as novel fingerprints can be published as (small) scripts and imported into different applications. For further explanations and generation of custom fingerprints, we refer to the documentation (\url{https://madas.readthedocs.io}).

Currently, \MADAS\ supports the spectral fingerprint used in Refs.~\cite{Kuban2022a, Kuban2022b} (\texttt{DOSFingerprint}), the PTE (Peridoc Table of Elements) descriptor of Ref. \cite{Kuban2022a} (\texttt{PTEFingerprint}), as well as fingerprint for scalar properties (\texttt{PROPFingerprint}) and a test fingerprint (\texttt{DUMMYFingerprint}) for demonstration and testing purposes.

\begin{figure}[htb]
    \centering
    \includegraphics[width = 0.8\textwidth]{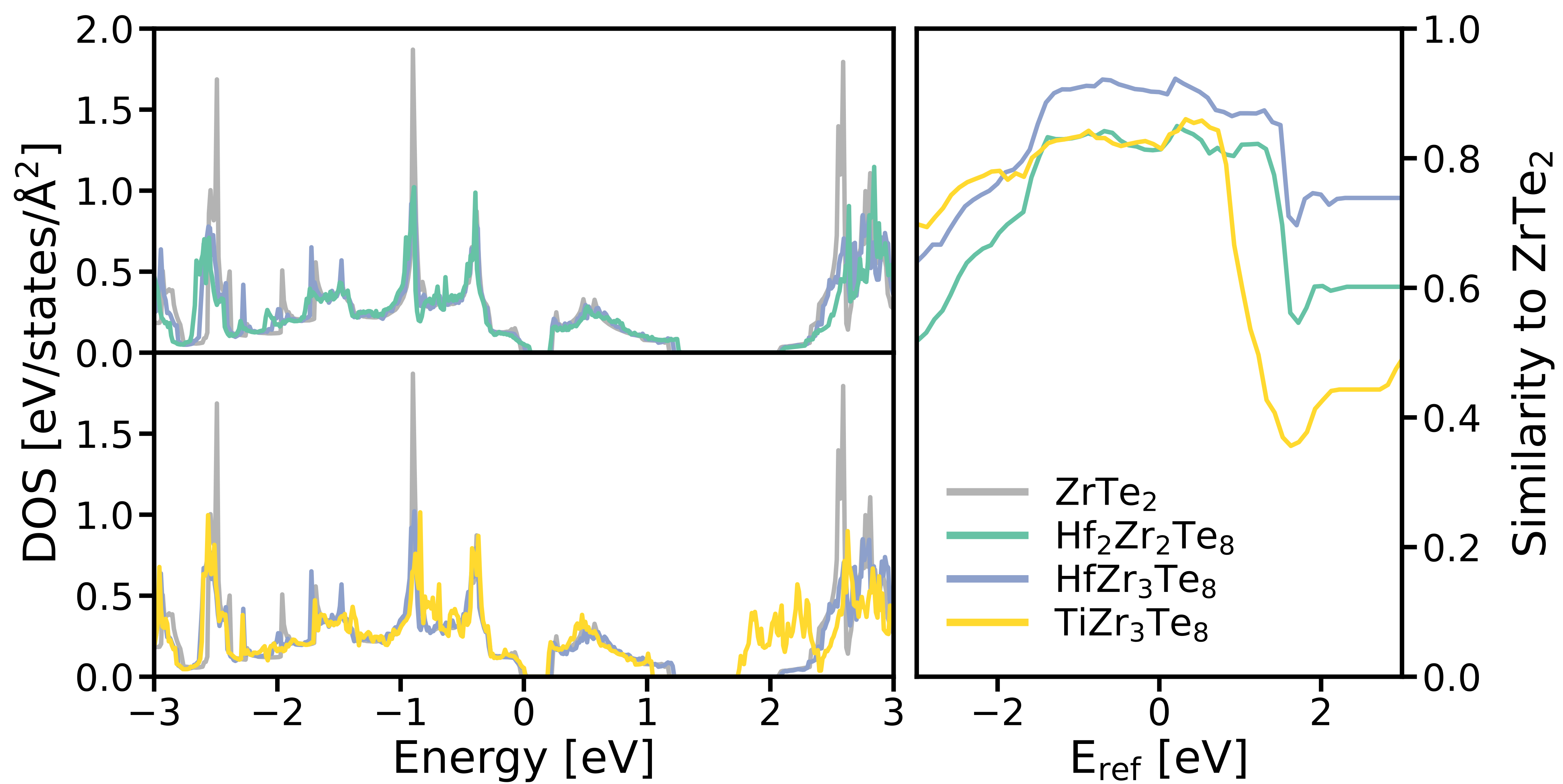}
    \caption{Materials most similar to ZrTe$_2$: The top (bottom) left panel shows the DOS of the most similar materials when considering the conduction (valance) bands as the feature region. The right panel shows the energy-resolved similarity in a region around the band gap.}
    \label{fig:fingerprint_comparison}
\end{figure}

The modular structure of \MADAS\ allows for seamlessly calculating fingerprints that can be used to study different aspects of the data. To exemplify this, we use data from the C2DB \cite{Haastrup2018, Gjerding2021b} and visualize in Fig. \ref{fig:fingerprint_comparison} the materials most similar to ZrTe\fsub{2} using \texttt{DOSFingerprint}s with different parameterizations. Here, the DOS is discretized on a grid and represented as a binary-valued vector. The grid can be varied by various parameters to emphasize an energy region considered most relevant when comparing different spectra, \ie the {\it feature region}. These parameters include the reference energy $\mathrm{E}_\mathrm{ref}$ (\ie the location of the grid's center) and the width $w$ of the feature region, as well as an energy cutoff that controls the total width of the fingerprint. In this example, we show how the results depend on the chosen energy range: The most similar material to ZrTe\fsub{2} is HfZr\fsub{3}Te\fsub{8}, irrespective of whether we put the focus on the valence bands ($\mathrm{E}_\mathrm{ref}=-2$ eV, $w=4$ eV), or on the conduction bands ($\mathrm{E}_\mathrm{ref}=2$ eV, $w=4$ eV), resulting in similarity values of $S=0.83$ in both cases. The next most similar ones are Hf\fsub{2}Zr\fsub{2}Te\fsub{8} (top left panel) and TiZr\fsub{3}Te\fsub{8} (bottom left panel) for both choices of the energy range, with a similarity of $S=0.75$ and $S=0.76$, respectively. To demonstrate which energy regions have the highest impact on the similarity, we compute spectral fingerprints in narrower energy windows of $w=2$ eV, centered at a range of different reference energies, \ie between $-3$ and $3$ eV. We then calculate the similarities between ZrTe\fsub{2} and its most similar materials for each of the reference energies. The result is shown in the right panel of Fig.~\ref{fig:fingerprint_comparison}. HfZr\fsub{3}Te\fsub{8} is most similar to ZrTe\fsub{2} over almost the whole energy range with the exception of the lower valence bands ($E_{ref} < -1$ eV), where TiZr\fsub{3}Te\fsub{8} has a higher similarity score. In the upper valence bands ($E_{ref}  > 1$ eV), Hf\fsub{2}Zr\fsub{2}Te\fsub{8} has a higher similarity score than the former. Around the Fermi energy ($-1$ eV $\leq E \leq 1$ eV), their similarities to the reference material are almost identical. 

This kind of analysis can be used to quantify differences in scientific results, making the analysis of spectra machine readable, and improving trust in data by quantitative analysis.

\subsection{Similarity matrix}
\label{sec:similarity_matrix}
Given a set of fingerprints, the similarity relations between them can be calculated. The \textit{similarity matrix} contains all pairwise similarities between members of a dataset. With a symmetric similarity measure, \ie $S(A, B) = S(B, A)$ for any two fingerprints $A$ and $B$, the matrix is obviously symmetric. In \MADAS, calculation, manipulation, and storage of similarity matrices is implemented via a \texttt{SimilarityMatrix} class. 
\begin{figure}[!htb]
    \centering
    \includegraphics[width = 0.75\textwidth]{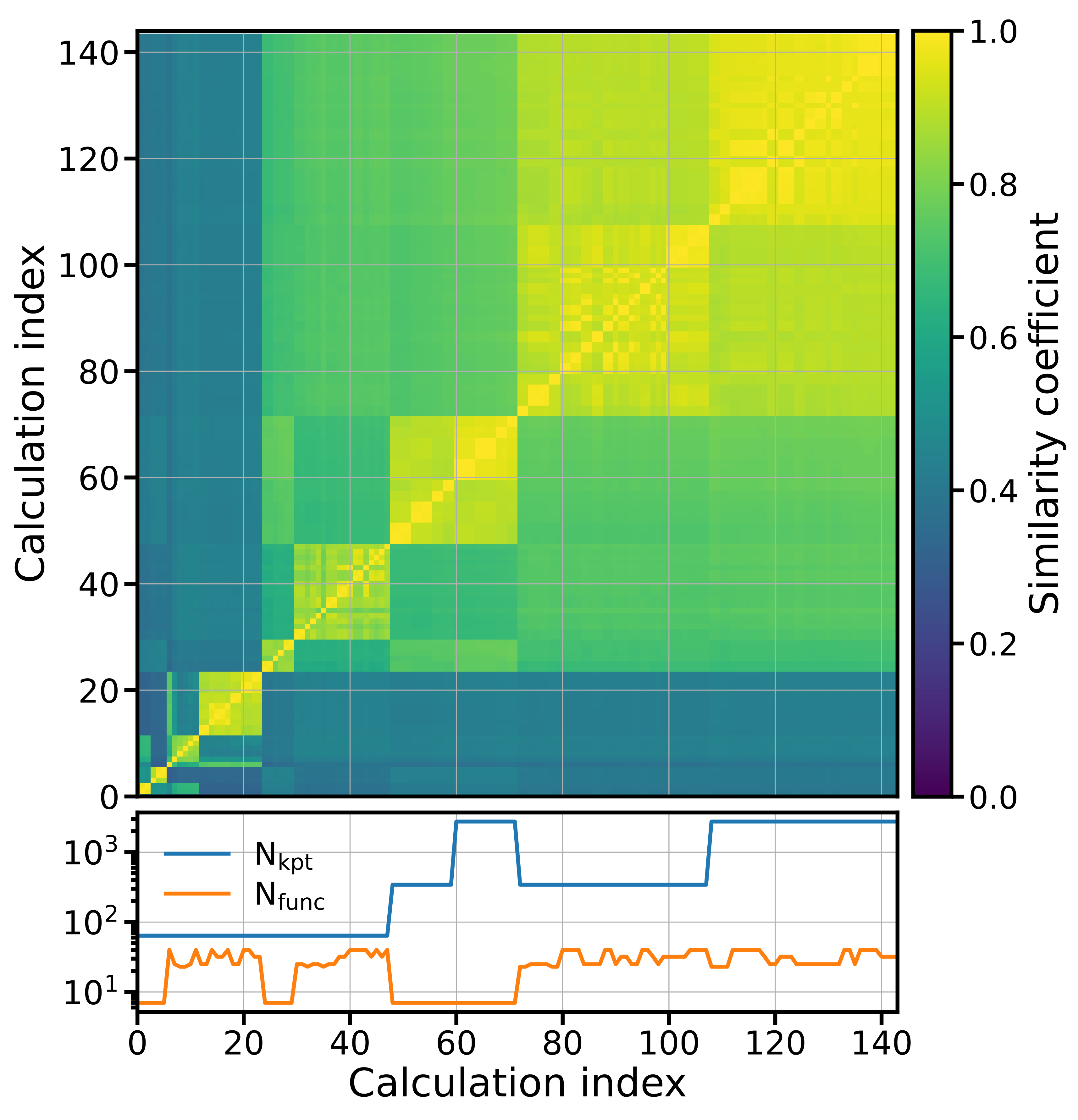}
    \caption{Similarity matrix of the DOS of AlGaO\fsub{3} from data obtained with different basis-set sizes and k-point sets. The calculation with the highest average similarity to the rest of the dataset has the highest index. The bottom panel displays the number of k-points, N$_\mathrm{kpt}$, and the number of basis functions, N$_\mathrm{func}$, that were used for the ground-state calculations. For the calculation of the DOS, 9 times more k-points were used. The color code indicates the similarity coefficient, ranging from 0 (dark blue) to 1 (yellow).}
    \label{fig:similarity_matrix}
\end{figure}

Figure~\ref{fig:similarity_matrix} shows a similarity matrix of the electronic DOS of AlGaO\fsub{3}, obtained with the DFT code FHI-aims. These calculations are part of a dataset used to study the numerical quality of DFT calculations \cite{Carbogno2022}. The subset used here has identical unit cell volumes, while several other parameters vary, such as the number of k-points used for Brillouin zone sampling, the basis set size, the relativistic treatment, and the exchange-correlation functional. In the figure, we sort the matrix rows and columns by their mean similarity to the rest of the dataset, \ie the calculation that is on average most similar to all other calculations has the highest calculation index $i_\mathrm{calc}$. Below the matrix, we show the number of k-points (blue) and the number of basis functions per atom (orange). The matrix exhibits a clear block structure, \ie there are subsets of calculations, whose members are more similar to other members of the set than to other calculations. There is a clear correlation between the computational parameter and the average similarity to the rest of the dataset. The calculations with lowest indices ($i_\mathrm{calc}\leq23$) are especially dissimilar, with average similarities $\bar{\mathrm{S}} \leq 0.5$. We traced this back to artifacts in the DOS which appear when the scalar ZORA approximation is used for the relativistic treatment of core electrons in combination with too few $k$-points. When more $k$-points are used, these artifacts disappear. The next block in the matrix consists of calculations which have different combinations of low numbers of $k$-points and/or basis functions. The last block $i_\mathrm{calc}>71$) with high average similarity has both sufficient $k$-points and basis-set sizes. We note in passing that analyzing similarity measures in the form illustrated here is generally only meaningful if a sufficiently large number of calculations are available. 

Depending on the type of fingerprint that is used, calculating similarity matrices can be computationally demanding, due to the quadratic scaling, $\mathcal{O}$($N^2)$, in the set size $N$. Therefore, we provide tools to optimize this task by considering individual entries, by parallelization, and/or by avoiding repetition of expensive calculations through storing results. Very large sets of fingerprints (about $\mathcal{O}$(10$^6)$) or computationally demanding similarity measures, may require to execute this task on a high-performance-computing (HPC) cluster. Since the entries of a similarity matrix are independent of each other, they can be computed in independent blocks. We provide an implementation of this functionality in a \texttt{BatchedSimilarityMatrix} class.

As an example of a use case for this implementation, we refer to the NOMAD Encyclopedia\cite{Draxl2019} (\url{https://nomad-lab.eu/prod/rae/encyclopedia}), which features a list of materials with the most similar electronic density of states (DOS), for the majority of its entries. To obtain this information, we computed the full similarity matrix for all $\sim 1.8$ million materials for which a DOS was available. This was only feasible by massive parallelization, as supported by the \texttt{BatchedSimilarityMatrix}.

\subsection{AI tools} \label{sec:AItools}
\MADAS\ can be used with a variety of artificial intelligence (AI) tools, including supervised and unsupervised learning. To achieve this, the functions and methods defined in \MADAS\ are designed to be compatible with the API of \texttt{scikit-learn}\cite{Pedregosa2011}. For example, to find sets of materials that are similar to each other, a similarity matrix can be used as input for clustering algorithms\cite{Kuban2022a}. Different to similarity searches (see Sec. \ref{sec:similarity_matrix}), clustering, as an unsupervised learning task, reveals global features of the dataset, by finding all sets of similar materials simultaneously.

\begin{figure}[!htb]
    \centering
    \includegraphics[width = \textwidth]{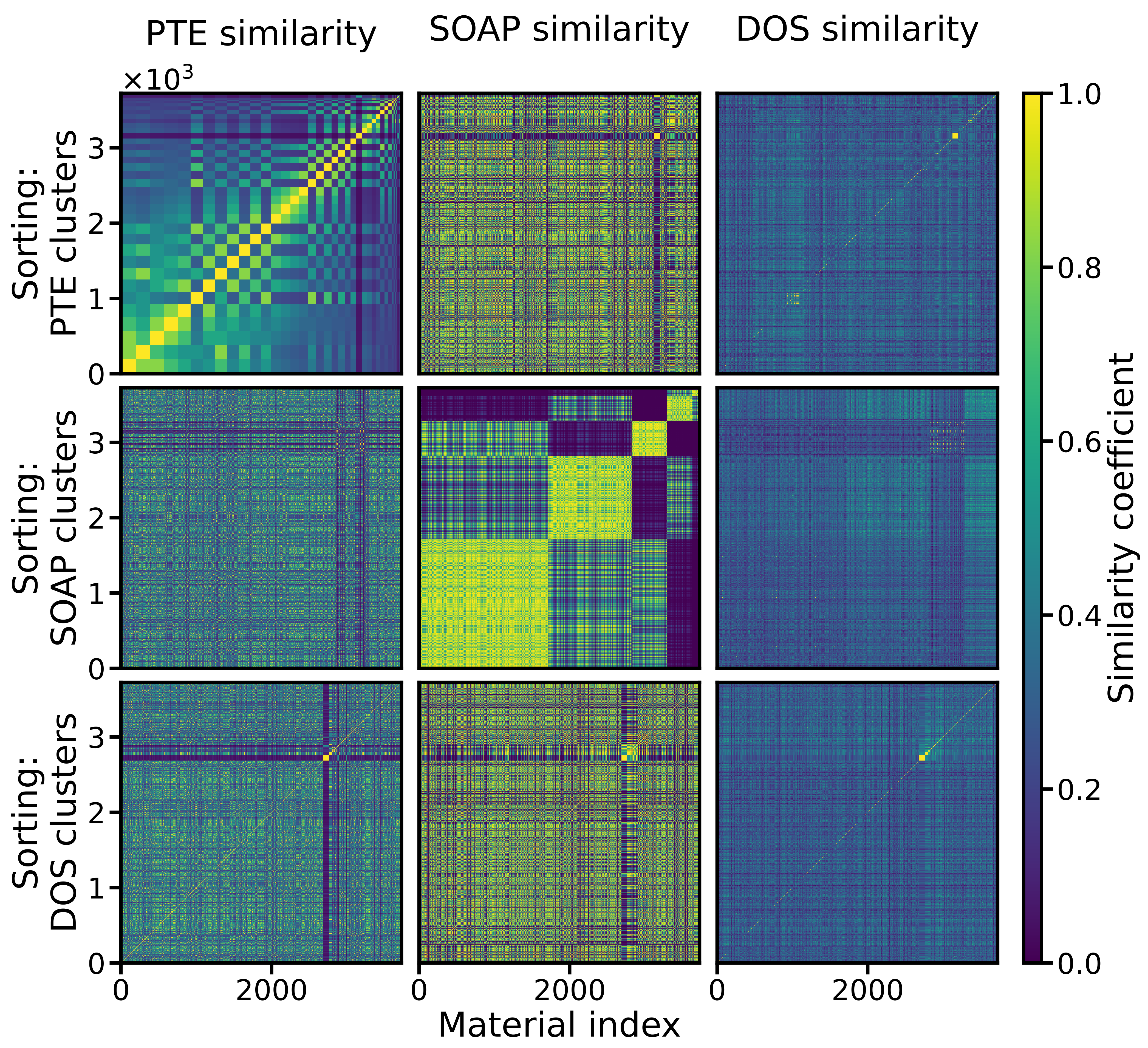}
    \caption{Similarity matrices computed for $\sim3800$ cubic perovskites. The columns of the grid correspond to the PTE (left), SOAP (middle), and DOS (right) fingerprints. In the top (middle, bottom) row of the grid, the materials in the similarity matrices are sorted such that they represent clusters that are found using the PTE (SOAP, DOS) similarity. The similarity coefficient is color coded, ranging from 0 (dark blue) to 1 (yellow).}
    \label{fig:AITools}
\end{figure}

In the following, we show how clustering can be used to analyze correlations between material properties. For this example, we have downloaded the crystal structures and electronic DOS for a dataset of 3847 cubic perovskites, which stem from the AFLOWLib database\cite{Curtarolo2012lib} and are also accessible through NOMAD\cite{Draxl2018}. We subsequently calculate PTE, DOS, and SOAP fingerprints (see Section \ref{sec:fingerprinting}) and the respective similarity matrices. The SOAP fingerprints reflect the atomic structure, but do not distinguish between atomic species. They are generated using \MADAS ' interfaces to \texttt{ASE} and \texttt{dscribe}: We obtain the atomic structure from \texttt{Material} objects (see Sec. \ref{sec:components}) as \texttt{ASE} \texttt{Atoms} objects and set all atomic species to the same element. Then we use the SOAP generator from \texttt{dscribe} to obtain the descriptor values, averaged over atom sites. As a similarity metric we use the pairwise Gaussian kernel of \texttt{scikit-learn}. We then cluster the similarity matrices using the threshold clustering method introduced in Ref.~\cite{Kuban2022a}. It can be used to find compact clusters where the similarity between cluster members is guaranteed to be larger than $2 S_\mathrm{thres} - 1$, where $S_\mathrm{thres}$ is the threshold used for clustering. We used a threshold of $S_\mathrm{thres}=1$ for the PTE matrix, \ie all cluster members have an identical PTE descriptor, and a threshold of $S_\mathrm{thres}=0.75$ for the DOS and SOAP matrices.

\MADAS\ provides the wrapper class \texttt{SimilarityMatrixClusterer}, which simplifies the use of clustering algorithms, specifically, any clustering method that uses the naming conventions introduced in \texttt{scikit-learn}. Figure \ref{fig:AITools} shows the similarity matrices for the PTE (left column), SOAP (middle column), and DOS (right column) fingerprints, sorted by the results of the clustering process on the PTE matrix (top row), the SOAP matrix (center row), and the DOS matrix (bottom row). The PTE matrix sorted by PTE clusters (top left panel) shows a many clusters of similar size, indicating a homogeneous distribution of elements across the materials in this dataset. Given the high-throughput, combinatorial approach of the AFLOWlib database, this was to be expected. The DOS matrix sorted by DOS clusters (bottom right panel) shows a different picture: The majority of the materials are outliers (indices $< 2686$), \ie they don't have a similarity higher than $S=0.75$ to any other material. At higher indices, all materials are contained in clusters, as defined by the \texttt{SimilarityMatrixClusterer} class. However, we find only few large clusters, demonstrating the chemical diversity in this dataset. The SOAP matrix sorted by SOAP clusters (middle panel) shows 10 large clusters of different sizes, where the largest one has almost 2000 members. Since the descriptor that we use in this example does not distinguish between atomic species and the data set consists only of cubic perovskites, the cluster formation is likely to be related to the cell volumes. 

Even more interesting are the off-diagonal elements of Fig. \ref{sec:AItools}: Looking at the PTE similarity matrix sorted by DOS clusters (bottom left), the we see that the largest cluster, which contains 75 materials (indices 2686 to 2761), is also visible in the PTE matrix. Moreover, we also find it in the SOAP matrix. Upon closer inspection, it turns out that these are all calculations of BPBa$_3$ \cite{FootnoteDuplictatePerovskites}. This demonstrates the usefulness of our approach for detecting duplicates. The DOS similarity matrix sorted by PTE clusters (top right) reveals a slight correlation with the PTE descriptor, \ie a block-like structure appearing in the DOS similarity matrix. One might argue that materials with the same composition, and thus PTE descriptor, are statistically more likely to have a similar DOS. However, similarity in the electronic structure depends on many different factors, so the PTE descriptor certainly does not contain enough information to explain the DOS similarity. In the DOS matrix sorted by SOAP clusters (middle right), we see that the largest SOAP cluster shows no correlation with the DOS. However, the smaller clusters (cluster index $>1717$) show a slightly higher than average similarity. Aside from the duplicate entries discussed above, the PTE and SOAP matrices do not appear to be correlated. This aligns with our expectations, because they are agnostic of each other by construction.

The above analysis can be performed using \MADAS\ in a few lines of code and can be found in the GitHub repository accompanying this paper (\url{https://github.com/kubanmar/madas-examples}). Moreover, to perform a similar analysis on another dataset, only the data query needs to be modified, resulting in a flexible and reusable workflow. We note that the analysis in Ref. \cite{Kuban2022a}, also employed the unsupervised learning interface of \MADAS.

We note also that supervised learning can be performed by either extracting descriptor values from \texttt{Fingerprint} objects, or by using similarity matrices as input. For the former, they can be retrieved by using the \texttt{Fingerprint().data[feature\_name]} attribute. Properties that are stored in the database can also be retrieved as a machine-learning-library friendly \texttt{pandas} \texttt{DataFrame} objects from the \texttt{MaterialsDatabase} (see Sec. \ref{sec:database}). Similarity matrices can be used as kernel matrices for kernel-based machine-learning algorithms. To do so, the values stored in a \texttt{SimilarityMatrix} can be accessed by its \texttt{matrix} attribute.

\section{Discussion and Outlook}
\label{sec:discussion}
With \MADAS, we present a Python-based framework to support all steps of similarity analysis, including the collection and storage of data, the development and computation of fingerprints, the calculation of large similarity matrices, and data analytics. At the same time, \MADAS\ is written in a modular way, allowing to customize it to the application at hand, and to use only the parts of the code that are necessary for that particular task. The benefits of using \MADAS\ lie not primarily in performance, but rather in flexibility and efficiency in prototyping and scripting (via an object model that favors re-usability), error-tolerance through customizable exception handling, focus on data provenance by logging, and integration with well-established libraries.

We have demonstrated its use for managing data and comparing calculations across different external data sources, have shown how spectral fingerprints can be used and adapted to quantify local and global similarities of the electronic structure of materials, and have exemplified how similarity matrices can be used to group and rank calculations performed with different numerical settings.

Similarity searches are a well-known technique in molecular chemistry and drug discovery \cite{Maggiora2014, Willett1998}. A common application is to scan a database of existing and hypothetical molecules for those that resemble a specific structural pattern that is assumed to correlate with favorable properties of a reference molecule. To do so, it is assumed that the presence or absence of molecular features, encoded in a fingerprint, correlates with the properties of the reference in terms of a so-called quantitative structure-property relationship (QSPR). That this is true in many cases is confirmed by the successes and continuous development of this technique. For materials, however, we find a different picture: The electronic structure (and therefore many derived properties of interest) is not necessarily determined by the local atomic structure, but reflects the intricate non-local many-body nature of extended systems. This situation asks for the development of novel, advanced fingerprints and techniques that can help to scan the materials space for interesting compounds.

In the future, we will extend \MADAS\ with more tools to support (semi)automatic data analysis, outlier detection, and focus on data-quality assessment. The longevity of the code is supported by its rich documentation (\url{https://madas.readthedocs.io}) and modular design. This opens up development to the community and enables the code to grow with its user base.

\section{Data availability}
The identifier for the data stemming from NOMAD, AFLOWlib, the Materials Project, and OQMD used in Sec. \ref{sec:Introduction}, Sec. \ref{sec:similarity_matrix}, and Sec. \ref{sec:AItools}, respectively, can be found in the GitHub repository for this manuscript (\url{https://github.com/kubanmar/madas-examples}). The data used in Sec.  \ref{sec:fingerprinting} are available at \url{https://github.com/kubanmar/dos-fingerprints-data}.

\section{Code availability}
\MADAS\ is released as open source under the Apache 2.0 license and available at the public Github repository \url{https://github.com/kubanmar/madas}. The documentation source code can be found at \url{https://github.com/kubanmar/madas-docs}. All code required to reproduce the results in this manuscript is available at \url{https://github.com/kubanmar/madas-examples}.

\section{Author contributions}
M.K. wrote the code and documentation and analyzed the results. S.R. and C.D. contributed ideas and discussed and reviewed all parts of the work. All authors participated in the writing of the manuscript.

\section{Acknowledgements}
This work received funding from the German Research Foundation (DFG) through the CRC 1404 (FONDA), project 414984028 and the NFDI consortium FAIRmat, project 460197019. We thank \v{S}imon Gabaj for testing the code and Lauri Himanen for valuable feedback to the manuscript.

\bibliographystyle{unsrt}
\bibliography{main}

\end{document}